\documentclass[aps,prd,twocolumn,superscriptaddress]{revtex4}
\usepackage{epsfig,epsf}
\usepackage{amsmath}
\usepackage{amsthm}
\usepackage{amsfonts}
\usepackage{amssymb}
\usepackage{dsfont}
\usepackage{multirow}
\usepackage{appendix}
\usepackage{slashed}
\usepackage[active]{srcltx}
\usepackage{psfrag}

\begin{document}

\title{Doubly charged vector tetraquark $Z_{\mathrm{V}}^{++}=[cu][\overline{s}\overline{d}]$}
\date{\today}
\author{S.~S.~Agaev}
\affiliation{Institute for Physical Problems, Baku State University, Az--1148 Baku,
Azerbaijan}
\author{K.~Azizi}
\affiliation{Department of Physics, University of Tehran, North Karegar Avenue, Tehran
14395-547, Iran}
\affiliation{Department of Physics, Do\v{g}u\c{s} University, Acibadem-Kadik\"{o}y, 34722
Istanbul, Turkey}
\author{H.~Sundu}
\affiliation{Department of Physics, Kocaeli University, 41380 Izmit, Turkey}

\begin{abstract}
We explore properties of the doubly charged vector tetraquark $Z_{\mathrm{V}%
}^{++}=[cu][\overline{s}\overline{d}]$ built of four quarks of different
flavors using the QCD sum rule methods. The mass and current coupling of $Z_{%
\mathrm{V}}^{++}$ are computed in the framework of the QCD two-point sum
rule approach by taking into account quark, gluon and mixed vacuum
condensates up to dimension $10$. The full width of this tetraquark is
saturated by $S$-wave $Z_{\mathrm{V}}^{++} \to \pi ^{+}D_{s1}(2460)^{+},\
\rho^{+}D_{s0}^{\ast }(2317)^{+}$, and $P$-wave $Z_{\mathrm{V}}^{++} \to \pi
^{+}D_{s}^{+},\ K^{+}D^{+}$ decays. Strong couplings required to find
partial widths of aforementioned decays are calculated in the context of the
QCD light-cone sum rule method and a soft-meson approximation. Our
predictions for the mass $m=(3515 \pm 125)~\mathrm{MeV}$ and full width $%
\Gamma _{\mathrm{full}}=156_{-30}^{+56}~\mathrm{MeV}$ of $Z_{\mathrm{V}%
}^{++} $ are useful to search for this exotic meson in various processes.
Recently, the LHCb collaboration discovered neutral states $X_{0(1)}(2900)$
as resonance-like peaks in $D^{-}K^{+}$ invariant mass distribution in the
decay $B^{+} \to D^{+}D^{-}K^{+}$. We argue that mass distribution of $%
D^{+}K^{+}$ mesons in the same $B$ decay can be used to observe the doubly
charged scalar $Z_{\mathrm{S}}^{++}$ and vector $Z_{\mathrm{V}}^{++}$
tetraquarks.
\end{abstract}

\maketitle



\section{Introduction}

\label{sec:Int} 

Hadrons with unusual quantum numbers and/or multi quark-gluon contents
attracted interest of researches starting from first days of the
quark-parton model and Quantum Chromodynamics (QCD). It is known that
conventional hadrons are composed of quark-antiquark pairs $q\overline{q}$,
or made of three valence quarks $qq^{\prime }q^{\prime \prime }$. Masses and
quantum parameters of ordinary mesons and baryons agree with predictions
obtained in this scheme and can be calculated using standard methods of high
energy physics. But fundamental principles of QCD do not forbid existence of
particles built of four, five, etc. quarks, containing valence gluons or
composed of exclusively valence gluons. Such hadrons may be produced in
decays of $B$ meson, in the electron-positron and proton-antiproton
annihilations $e^{+}e^{-}$ and $\overline{p}p$ , in the double charmonium
production processes, in the two-photon fusion and $pp$ collisions.

The concept of multiquark hadrons was applied by R.~Jaffe to explain a mass
hierarchy inside the lowest scalar multiplet \cite{Jaffe:1976ig}. In
accordance with this assumption the nonet of the light scalars are
four-quark states $q^{2}\overline{q}^{2}$, which explain exprementally
observed features of these particles. Another interesting result about
multiquark hadrons was obtained also by R.~Jaffe \cite{Jaffe:1976yi}. Thus,
he considered six-quark states composed of only light quarks, and using the
MIT quark-bag model calculated parameters of flavor-singlet and\ neutral
bound state $uuddss$ ($\mathrm{H}$-dibaryon) with isospin $I=0$ and
spin-parity $J^{\mathrm{P}}=0^{+}$. If exist, this double-strange structure
would be stable against strong decays, and have mean lifetime $\tau \approx
10^{-10}\mathrm{s}$, which is considerably longer than that of conventional
mesons.

Properties of exotic hadrons were investigated in the context of
QCD-inspired nonperturbative methods already at eighties of the last
century. One of such effective methods is the QCD sum rule approach \cite%
{Shifman:1978bx,Shifman:1978by}, which was employed to perform quantitative
analyses and calculate masses and other parameters of the glueballs, hybrid $%
q\overline{q}g$ and four-quark mesons (tetraquarks) \cite%
{Shifman:1978bx,Balitsky:1982ps,Govaerts:1984hc,Govaerts:1985fx,Balitsky:1986hf,Braun:1985ah,Braun:1988kv}%
. Unfortunately, achievements obtained at early stages of theoretical
studies were not accompanied by reliable experimental measurements connected
mainly with difficulties in detecting heavy resonances. As a result,
existence of exotic hadrons was not then certainly established.

This situation changed starting from the Belle's discovery of the
charmoniumlike resonance $X(3872)$ \cite{Choi:2003ue}, which was confirmed
later by other collaborations \cite%
{Abazov:2004kp,Acosta:2003zx,Aubert:2004ns}. During followed years different
experimental groups collected information about masses, widths, and quantum
numbers of numerous heavy resonances, which may be considered as four-quark
exotic mesons. There were attempts to describe new charmoniumlike states as
excitations of the ordinary charmonium $c\overline{c}$: It is worth noting
that some of them really allows such interpretation. But the bulk of
available experimental data cannot be included into this scheme. These
resonances may be considered in a tetraquark model, which treats them either
as two-meson molecules or diquark-antidiquark states.

Observation of the charged resonances $Z^{\pm }(4430)$ and $Z_{c}^{\pm
}(3900)$ had important impact on the physics of multiquark hadrons, because
they could not be interpreted as neutral charmonia and became real
candidates to tetraquarks \cite{Choi:2007wga,Ablikim:2013mio}. Exotic mesons
containing four quarks of different flavors also differ from charmoniumlike
states and are promising candidates to genuine four-quark mesons. Analyses
of such structures were inspired by information of the $D0$ collaboration
about evidence for the state $X(5568)$ composed presumably of four different
quarks \cite{D0:2016mwd}. Later, other collaborations could not confirm
existence of this state \cite{Aaij:2016iev,CMS:2016fvl}, but knowledge
gained due to theoretical studies of $X(5568)$, methods and schemes
elaborated during this process played an important role in our understanding
of exotic mesons. The resonances which are candidates to four-quark exotic
mesons now form the wide family of $XYZ$ states. Detailed information on $%
XYZ $ \ particles including a history of the problem, as well as
experimental results and theoretical achievements of last years are
collected in numerous interesting review articles \cite%
{Jaffe:2004ph,Swanson:2006st,Klempt:2007cp,Chen:2016qju,
Chen:2016spr,Esposito:2016noz,Ali:2017jda,Olsen:2017bmm,
Albuquerque:2018jkn,Brambilla:2019esw,Agaev:2020zad}.

Recently the LHCb collaboration announced about new resonance-like
structures $X_{0(1)}(2900)$ observed in the process $B^{+}\rightarrow
D^{+}D^{-}K^{+}$ \cite{Aaij:2020hon,Aaij:2020ypa}. They were seen in the $%
D^{-}K^{+}$ mass distribution, and can be considered as first strong
evidence for exotic mesons composed of four quarks of different flavors.
Indeed, from decay modes of these states, it is clear that they contain
valence quarks $ud\overline{c}\overline{s}$. But one should bear in mind
that $X_{0(1)}(2900)$ may have alternative origin, and appear due to
rescattering diagrams $\chi _{c1}D^{\ast -}K^{\ast +}$ and $D_{sJ}\overline{D%
}_{1}^{0}K^{0}$ in the decay $B^{+}\rightarrow D^{+}D^{-}K^{+}$ \cite%
{Liu:2020orv}.

The LHCb determined masses and spin-parities of these resonances, that were
used in various models to explain internal organizations of $X_{0(1)}(2900)$%
. As usual, they were interpreted as hadronic molecules, diquark-antidiquark
states, and rescattering effects (see, Refs.\ \cite%
{Agaev:2020nrc,Agaev:2021knl} and references therein). In our articles \cite%
{Agaev:2020nrc,Agaev:2021knl}, we treated $X_{0}(2900)$ and $X_{1}(2900)$ as
a scalar hadronic molecule $\overline{D}^{\ast 0}K^{\ast 0}$ and
diquark-antidiquark vector states $[ud][\overline{c}\overline{s}]$,
respectively. Predictions for their masses and widths extracted from sum
rule analyses seem confirm assumptions made on their structures.

The resonances $X_{0(1)}(2900)$ emerge in intermediate phase of the decay
chain $B^{+}\rightarrow D^{+}X\rightarrow D^{+}D^{-}K^{+}$, and are neutral
states. These processes occur due to color-favored and color-suppressed
transformations of the $B^{+}$ meson \cite{Burns:2020xne}. But weak decays
of $B^{+}$ with the same topologies may trigger also processes $%
B^{+}\rightarrow D^{-}Z^{++}\rightarrow D^{-}D^{+}K^{+}$, where $Z^{++}$ is
a doubly charged exotic meson with quark content $cu\overline{s}\overline{d}$%
. In our view, experimental data collected by LHCb about weak decays of $B$
meson may be used to uncover doubly charged tetraquarks $Z^{++}=[cu][%
\overline{s}\overline{d}]$ with different spin-parities. In fact, the scalar
and vector particles $Z_{\mathrm{S}}^{++}$ and $Z_{\mathrm{V}}^{++}$ may
appear as resonances in the $D^{+}K^{+}$ mass distribution, and provide
valuable information on new exotic mesons.

Let us note that doubly charged diquark-antidiquarks already attracted
interests of physicists, and some of them was studied in a rather detailed
form. Thus, spectroscopic parameters and strong decays of pseudoscalar
tetraquarks $cc\overline{s}\overline{s}$ and $cc\overline{d}\overline{s}$
were calculated in Ref.\ \cite{Agaev:2018vag}. Doubly charged scalar,
pseudoscalar and axial-vector states $[sd][\overline{u}\overline{c}]$ were
considered in our article \cite{Agaev:2017oay}. The tetraquarks $Z^{++}$ are
positively charged counterparts of states $[sd][\overline{u}\overline{c}]$
and have the same masses and decay widths. Therefore, one can safely use
their parameters to study exotic mesons $Z^{++}$.

For completeness and following analysis, we provide masses and widths of
various tetraquarks $Z^{++}$ using, for these purposes, our results from
Ref.\ \cite{Agaev:2017oay}
\begin{eqnarray}
m_{\mathrm{Z}_{\mathrm{S}}} &=&2628_{-153}^{+166}~\mathrm{MeV},\ \Gamma _{%
\mathrm{Z}_{\mathrm{S}}}=(66.89\pm 15.11)~\mathrm{MeV,}  \notag \\
m_{\mathrm{Z}_{\mathrm{PS}}} &=&2719_{-156}^{+144}~\mathrm{MeV},\ \Gamma _{%
\mathrm{Z}_{\mathrm{PS}}}=(38.1\pm 7.1)~\mathrm{MeV,}  \notag \\
m_{\mathrm{Z}_{\mathrm{AV}}} &=&2826_{-157}^{+134}~\mathrm{MeV},\ \Gamma _{%
\mathrm{Z}_{\mathrm{AV}}}=(47.3\pm 11.1)~\mathrm{MeV.}  \notag \\
&&  \label{eq:DCh}
\end{eqnarray}%
Here, subscripts $Z_{\mathrm{S}}$, $Z_{\mathrm{PS}}$ and $Z_{\mathrm{AV}}$
refer to the scalar, pseudoscalar and axial-vector $Z^{++}$, respectively.

As is seen, there are not predictions for parameters of the vector
tetraquark $Z_{\mathrm{V}}^{++}$, but its mass should be around or larger
than $m_{\mathrm{Z}_{\mathrm{AV}}}$. Using this preliminary estimate for the
mass of $Z_{\mathrm{V}}^{++}$ and $m_{\mathrm{Z}_{\mathrm{S}}}$ from Eq.\ (%
\ref{eq:DCh}) , we see that decays of $Z_{\mathrm{S}}^{++}$ and $Z_{\mathrm{V%
}}^{++}$ to ordinary mesons $D_{s}^{+}\pi ^{+}$ and $D^{+}K^{+}$ are
kinematically allowed processes, i.e., the tetraquarks $Z_{\mathrm{S}}^{++}$
and $Z_{\mathrm{V}}^{++}$ may be seen in $D^{+}K^{+}$ mass distribution in
the decay $B^{+}\rightarrow D^{-}D^{+}K^{+}$.

In the present article, we compute spectroscopic parameters and total width
of the vector tetraquark $Z_{\mathrm{V}}^{++}$ using various versions of the
QCD sum rule method. Our interest to this particle is twofold: First, it is
a fully open flavor tetraquark, and additionally bears two units of electric
charge. By theoretical exploration of $Z_{\mathrm{V}}^{++}$, we will
complete list of such states (\ref{eq:DCh}). Second reason is that, there
are opportunities to fix the tetraquarks $Z_{\mathrm{S}}^{++}$ and $Z_{%
\mathrm{V}}^{++}$ using existing or future LHCb data.

We evaluate the mass and current coupling of $Z_{\mathrm{V}}^{++}$ by
employing the QCD two-point sum rule approach \cite%
{Shifman:1978bx,Shifman:1978by}. In our analysis, we take into account
contributions of various vacuum condensates up to dimension $10$. Prediction
for the mass of $Z_{\mathrm{V}}^{++}$, as well as its quantum numbers $J^{%
\mathrm{P}}=1^{-}$ permit us to classify kinematically allowed decay modes
of \ this tetraquark. The mass and coupling of $Z_{\mathrm{V}}^{++}$ are
also input parameters necessary to calculate partial widths of the decays $%
Z_{\mathrm{V}}^{++}\rightarrow D_{s1}(2460)^{+}\pi ^{+},\ D_{s0}^{\ast
}(2317)^{+}\rho ^{+}$, $\ D^{+}K^{+}$, and $D_{s}^{+}\pi ^{+}$. To this end,
we explore vertices of $Z_{\mathrm{V}}^{++}$ with ordinary mesons (for
example, $Z_{\mathrm{V}}^{++}D^{+}K^{+}$), and find corresponding strong
couplings. Relevant investigations are carried out using the QCD light-cone
sum rule (LCSR) method, which is one of effective nonperturbative tools to
study conventional hadrons \cite{Balitsky:1989ry}. In the case of
tetraquark-ordinary meson vertices the standard methods of LCSR have to be
applied in conjunction with a soft-meson approximation \cite%
{Belyaev:1994zk,Ioffe:1983ju}. For analysis of the exotic mesons the
light-cone sum rule method and soft approximation was suggested in Ref.\
\cite{Agaev:2016dev}, and used to investigate decays of various tetraquarks
\cite{Agaev:2020zad}.

This paper is structured in the following manner: Section \ref{sec:Masses}
is devoted to calculations of the mass and coupling of the vector tetraquark
$Z_{\mathrm{V}}^{++}=[cu][\overline{s}\overline{d}]$. In Section \ref%
{sec:Decays}, we compute strong couplings in relevant tetraquark-meson
vertices, and evaluate partial widths of $Z_{\mathrm{V}}^{++}$ decays. The
full width of $Z_{\mathrm{V}}^{++}$ is found in this section as well.
Section \ref{sec:Disc} is reserved for our conclusions and final notes.


\section{The spectroscopic parameters of $Z_{\mathrm{V}}$}

\label{sec:Masses}
The mass $m$ and current coupling $f$ are important parameters of the vector
tetraquark $Z_{\mathrm{V}}^{++}=[cu][\overline{s}\overline{d}]$ (in what
follows, we omit superscripts denoting charges of the tetraquark and various
mesons). We use the QCD two-point sum rule method to evaluate values of
these parameters.

We begin calculations from analysis of the two-point correlation function
\begin{equation}
\Pi _{\mu \nu }(p)=i\int d^{4}xe^{ipx}\langle 0|\mathcal{T}\{J_{\mu
}(x)J_{\nu }^{\dag }(0)\}|0\rangle ,  \label{eq:CF1}
\end{equation}%
where $J_{\mu }(x)$ is the interpolating current for $Z_{\mathrm{V}}$, and $%
\mathcal{T}$ \ denotes the time-ordered product of two currents. The vector
tetraquark $Z_{\mathrm{V}}$ can be modeled using a scalar diquark $uC\gamma
_{5}c$ and vector antidiquark $\overline{s}\gamma _{\mu }\gamma _{5}C%
\overline{d}$. It is known that color-antitriplet diquarks and color-triplet
antidiquarks are most stable two-quark structures \cite{Jaffe:2004ph}.
Therefore, we chose as building blokes of $Z_{\mathrm{V}}$ structures $%
\varepsilon ^{abc}u_{b}C\gamma _{5}c_{c}$ and $\varepsilon ^{amn}\overline{s}%
_{m}\gamma _{\mu }\gamma _{5}C\overline{d}_{n}^{T}$, which belongs to $[%
\overline{\mathbf{3}}_{c}]$ and $[\mathbf{3}_{c}]$ representations of the
color group $SU_{c}(3)$, respectively. Then, the colorless interpolating
current $J_{\mu }(x)$ takes the form
\begin{equation}
J_{\mu }(x)=\varepsilon \widetilde{\varepsilon }[u_{b}^{T}(x)C\gamma
_{5}c_{c}(x)][\overline{s}_{m}(x)\gamma _{\mu }\gamma _{5}C\overline{d}%
_{n}^{T}(x)],  \label{eq:CR1}
\end{equation}%
where $\varepsilon \widetilde{\varepsilon }=\varepsilon ^{abc}\varepsilon
^{amn}$, and $a$, $b$, $c$, $m$ and $n$ denote quark colors. In Eq.\ (\ref%
{eq:CR1}) $u(x)$, $c(x)$, $s(x)$ and $d(x)$ are the quark fields, and $C$
stands for the charge-conjugation operator.

The sum rules for $m$ and $f$ can be derived by calculating $\Pi _{\mu \nu
}(p)$ in terms of the physical parameters of $Z_{\mathrm{V}}$, as well as in
the operator product expansion ($\mathrm{OPE}$) with certain accuracy.
Having equated these two expressions, applied the Borel transformation to
suppress contributions of higher resonances and continuum states, and
subtracted these contributions using the quark-hadron duality assumption
\cite{Shifman:1978bx,Shifman:1978by}, it is possible to get required sum
rules for the mass and coupling of the tetraquark $Z_{\mathrm{V}}$.

Expression of $\Pi _{\mu \nu }(p)$ in terms of parameters of $Z_{\mathrm{V}}$
is obtained by saturating the correlation function $\Pi _{\mu \nu }(p)$ with
a complete set of $J^{\mathrm{P}}=1^{-}$ states and performing in Eq.\ (\ref%
{eq:CF1}) integration over $x$
\begin{equation}
\Pi _{\mu \nu }^{\mathrm{Phys}}(p)=\frac{\langle 0|J_{\mu }|Z_{\mathrm{V}%
}\rangle \langle Z_{\mathrm{V}}|J_{\nu }^{\dagger }|0\rangle }{m^{2}-p^{2}}%
+\cdots ,  \label{eq:CF2}
\end{equation}%
where the dots stand for contributions of higher resonances and continuum
states.

To proceed, it is convenient to introduce the matrix element
\begin{equation}
\langle 0|J_{\mu }|Z_{\mathrm{V}}\rangle =fm\epsilon _{\mu },
\label{eq:MElem1}
\end{equation}%
where $\epsilon _{\mu }$ is the polarization vector of the tetraquark $Z_{%
\mathrm{V}}$. Then $\Pi _{\mu \nu }^{\mathrm{Phys}}(p)$ can be rewritten in
the following form
\begin{equation}
\Pi _{\mu \nu }^{\mathrm{Phys}}(p)=\frac{m^{2}f^{2}}{m^{2}-p^{2}}\left(
-g_{\mu \nu }+\frac{p_{\mu }p_{\nu }}{m^{2}}\right) +\cdots .
\label{eq:CorF1}
\end{equation}%
The two-component Lorentz structure in the parentheses in Eq.\ (\ref%
{eq:CorF1}) corresponds to a vector particle. The first component of this
structure is proportional to $g_{\mu \nu }$ and does not contain effects of
scalar particles: It is formed due to only vector states. Therefore, in our
studies we utilize the structure $\sim g_{\mu \nu }$, and label
corresponding invariant amplitude by $\Pi ^{\mathrm{Phys}}(p^{2})$.

Expression in Eq.\ (\ref{eq:CF2}) is obtained in the zero-width single-pole
approximation. In general, the correlation function $\Pi _{\mu \nu }(p)$
receives contributions also from two-meson reducible terms, because the
current $J_{\mu }$ couples not only to four-quark structures, but also to
two mesons with relevant quantum numbers \cite{Kondo:2004cr,Lee:2004xk}.
Interactions with mesons lying below the mass of the $Z_{\mathrm{V}}$
generate a finite width of the tetraquark $Z_{\mathrm{V}}$, and modify the
quark propagator in Eq.\ (\ref{eq:CF2})%
\begin{equation}
\frac{1}{m^{2}-p^{2}}\rightarrow \frac{1}{m^{2}-p^{2}-i\sqrt{p^{2}}\Gamma (p)%
}.
\end{equation}
The two-meson contributions can be either subtracted from the sum rules or
included into parameters of the pole term. In the case of the tetraquarks
the second approach is preferable and was applied in articles \cite%
{Wang:2015nwa,Agaev:2018vag,Sundu:2018nxt}. These effects, properly taken
into account in the sum rules, rescale the coupling $f$ \ leaving stable the
mass $m$ of the tetraquark. Detailed analyses proved that two-meson
contributions are small and can be neglected.

The QCD side of the sum rules $\Pi _{\mu \nu }^{\mathrm{OPE}}(p)$ is found
by inserting the interpolating current $J_\mu(x)$ into Eq.\ (\ref{eq:CF1}),
and contracting relevant quark fields
\begin{eqnarray}
&&\Pi _{\mu \nu }^{\mathrm{OPE}}(p)=i\int d^{4}xe^{ipx}\varepsilon
\widetilde{\varepsilon }\varepsilon ^{\prime }\widetilde{\varepsilon }%
^{\prime }\mathrm{Tr}\left[ \gamma _{5}\widetilde{S}_{u}^{bb^{\prime
}}(x)\gamma _{5}S_{c}^{cc^{\prime }}(x)\right]  \notag \\
&&\times \mathrm{Tr}\left[ \gamma _{\mu }\gamma _{5}\widetilde{S}%
_{d}^{n^{\prime }n}(-x)\gamma _{5}\gamma _{\nu }S_{s}^{m^{\prime }m}(-x)%
\right] ,  \label{eq:QCD1}
\end{eqnarray}%
where
\begin{equation}
\widetilde{S}_{c(q)}(x)=CS_{c(q)}^{T}(x)C,  \label{eq:Not}
\end{equation}%
with $S_{c}(x)$ and $S_{u(s,d)}(x)$ being the heavy $c$- and light $u(s,d)$%
-quark propagators, respectively (for explicit expressions, see Ref.\ \cite%
{Agaev:2020zad}). The invariant amplitude in $\Pi _{\mu \nu }^{\mathrm{OPE}%
}(p)$ corresponding to the structure $g_{\mu \nu }$ in our following
analysis will be denoted by $\Pi ^{\mathrm{OPE}}(p^{2})$.

Calculations performed in accordance with a scheme briefly explained above
yield
\begin{equation}
m^{2}=\frac{\Pi ^{\prime }(M^{2},s_{0})}{\Pi (M^{2},s_{0})},  \label{eq:Mass}
\end{equation}%
and
\begin{equation}
f^{2}=\frac{e^{m^{2}/M^{2}}}{m^{2}}\Pi (M^{2},s_{0}),  \label{eq:Coupl}
\end{equation}%
which are sum rules for $m$ and $f$, respectively. Here, $\Pi (M^{2},s_{0})$
is the Borel transformed and subtracted amplitude $\Pi ^{\mathrm{OPE}%
}(p^{2}) $, which depends on the Borel $M^{2}$ and continuum threshold $%
s_{0} $ parameters. In Eq.\ (\ref{eq:Mass}) $\Pi ^{\prime }(M^{2},s_{0})$ is
the derivative of the amplitude over $d/d(-1/M^{2})$.

It is clear that the amplitude $\Pi (M^{2},s_{0})$ is a key ingredient of
the obtained sum rules. Calculations of this function give the following
result
\begin{equation}
\Pi (M^{2},s_{0})=\int_{\mathcal{M}^{2}}^{s_{0}}ds\rho ^{\mathrm{OPE}%
}(s)e^{-s/M^{2}}+\Pi (M^{2}),  \label{eq:InvAmp}
\end{equation}%
where $\mathcal{M}=m_{c}+m_{s}$. Computations are carried out by taking into
account vacuum condensates up to dimension $10$. The amplitude $\Pi
(M^{2},s_{0})$ has two components: First of them is expressed using the
spectral density $\rho ^{\mathrm{OPE}}(s)$, which is computed as an
imaginary part of $\Pi _{\mu \nu }^{\mathrm{OPE}}(p)$. The Borel
transformation of another terms are found directly from $\Pi _{\mu \nu }^{%
\mathrm{OPE}}(p)$ and included into $\Pi (M^{2})$. The first component in
Eq.\ (\ref{eq:InvAmp}) contains a main part of the amplitude, whereas $\Pi
(M^{2})$ is formed from higher dimensional terms. Analytical expressions of $%
\rho ^{\mathrm{OPE}}(s)$ and $\Pi (M^{2})$, are rather lengthy, therefore we
do not write down them here.

The sum rules depend on vacuum condensates up to dimension $10$. The basic
condensates
\begin{eqnarray}
&&\langle \overline{q}q\rangle =-(0.24\pm 0.01)^{3}~\mathrm{GeV}^{3},\
\langle \overline{s}s\rangle =(0.8\pm 0.1)\langle \overline{q}q\rangle ,
\notag \\
&&\langle \overline{q}g_{s}\sigma Gq\rangle =m_{0}^{2}\langle \overline{q}%
q\rangle ,\ \langle \overline{s}g_{s}\sigma Gs\rangle =m_{0}^{2}\langle
\overline{s}s\rangle ,  \notag \\
&&m_{0}^{2}=(0.8\pm 0.2)~\mathrm{GeV}^{2}  \notag \\
&&\langle \frac{\alpha _{s}G^{2}}{\pi }\rangle =(0.012\pm 0.004)~\mathrm{GeV}%
^{4},  \notag \\
&&\langle g_{s}^{3}G^{3}\rangle =(0.57\pm 0.29)~\mathrm{GeV}^{6}.
\label{eq:Parameters}
\end{eqnarray}%
were estimated from analysis of numerous hadronic processes \cite%
{Shifman:1978bx,Shifman:1978by,Ioffe:2005ym,Narison:2015nxh}, and are known
and universal parameters. Higher dimensional condensates are factorized and
expressed in terms of basic ones: we assume that factorization does not lead
to large uncertainties. The masses of $s$ and $c\ $quarks are also among
input parameters, for which we use $m_{s}=93_{-5}^{+11}~\mathrm{MeV}$, and $%
m_{c}=1.27\pm 0.2~\mathrm{GeV}$, respectively.

The Borel and continuum threshold parameters $M^{2}$ and $s_{0}$ are
auxiliary quantities of computations and their choice\ is controlled by
constraints imposed on the pole contribution ($\mathrm{PC}$) and convergence
of $\mathrm{OPE}$, as well as by minimum sensitivity of the extracted
quantities to the Borel parameter $M^{2}$. Thus, the maximum allowed value
of $M^{2}$ should be fixed to meet the restriction on $\mathrm{PC}$%
\begin{equation}
\mathrm{PC}=\frac{\Pi (M^{2},s_{0})}{\Pi (M^{2},\infty )}.  \label{eq:PC}
\end{equation}%
The lower limit of the working window for the Borel parameter is found from
convergence of the operator product expansion, which is quantified by the
ratio%
\begin{equation}
R(M^{2})=\frac{\Pi ^{\mathrm{DimN}}(M^{2},s_{0})}{\Pi (M^{2},s_{0})}.
\label{eq:Convergence}
\end{equation}%
Here $\Pi ^{\mathrm{DimN}}(M^{2},s_{0})$ denotes contribution of the last
term (or a sum of last few terms) in the $\mathrm{OPE}$. In the present
work, at the maximum of $M^{2}$ we apply the constraint $\mathrm{PC>0.2}$,
which is typical for the multiquark hadrons. To guarantee the convergence of
the operator product expansion at minimum of $M^{2}$, we use a requirement $%
R(M^{2})\leq 0.01$.

Computations demonstrate that working regions for the parameters $M^{2}$ and
$s_{0}$
\begin{equation}
M^{2}\in \lbrack 4.5,6.5]\ \mathrm{GeV}^{2},\ s_{0}\in \lbrack 14,16]\
\mathrm{GeV}^{2},  \label{eq:Wind1}
\end{equation}%
satisfy aforementioned constraints. In fact, in these regions $\mathrm{PC}$
changes on average within limits
\begin{equation}
0.78\leq \mathrm{PC}\leq 0.28.
\end{equation}%
At the minimum $M^{2}=4.5~\mathrm{GeV}^{2}$, a contribution to $\Pi
(M^{2},s_{0})$ arising from a sum of last three terms in $\mathrm{OPE}$ does
not exceed $1\%$ of the full result. In fact, at $M^{2}=4.5~\mathrm{GeV}^{2}$
the ratio $R(M^{2})$ for $\mathrm{DimN=Dim(8+9+10)}$ is equal to $0.007$,
which proves convergence of the sum rules.

Central values of the mass $m$ and coupling $f$ are evaluated at the middle
point of regions (\ref{eq:Wind1}), in other words, at $M^{2}=5.5~\mathrm{GeV}%
^{2}$ and $s_{0}=15~\mathrm{GeV}^{2}$. At this point the pole contribution
is $\mathrm{PC}\approx 0.56$, which ensures the ground-state nature of $Z_{%
\mathrm{V}}$. Results for $m$ and $f$ read
\begin{eqnarray}
m &=&(3515~\pm 125)~\mathrm{MeV},  \notag \\
f &=&(5.24\pm 1.10)\times 10^{-3}~\mathrm{GeV}^{4}.  \label{eq:Result1}
\end{eqnarray}%
As it has been emphasized above, predictions extracted from sum rules should
not depend on a choice of $M^{2}$. In real analysis, however, there is a
residual dependence on this parameter. In Fig.\ \ref{fig:Mass}, we plot the
mass of the tetraquark $Z_{\mathrm{V}}$ for wide range of $M^{2}$ and $s_{0}$%
. It is seen, that only the region between two vertical lines in the left
panel can be considered as a relatively stable plateau, where parameters of $%
Z_{\mathrm{V}}$ are evaluated. The current coupling $f$ of the tetraquark $%
Z_{\mathrm{V}}$ is depicted in Fig.\ \ref{fig:Coupling} as functions of $%
M^{2}$ and $s_{0}$. In the case under discussion, one observes that $f$ is
almost stable in the explored range of the Borel parameter (\ref{eq:Wind1}).

\begin{widetext}

\begin{figure}[h!]
\begin{center}
\includegraphics[totalheight=6cm,width=8cm]{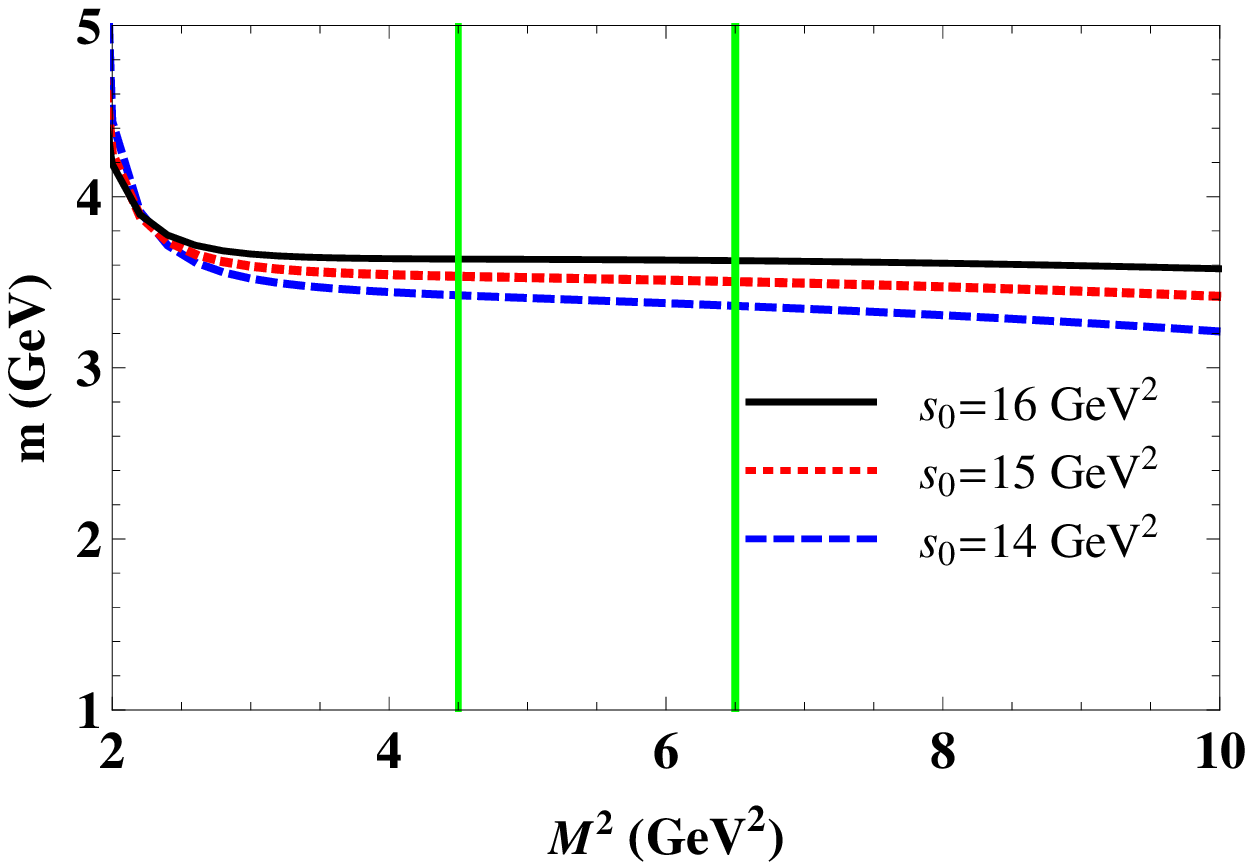}
\includegraphics[totalheight=6cm,width=8cm]{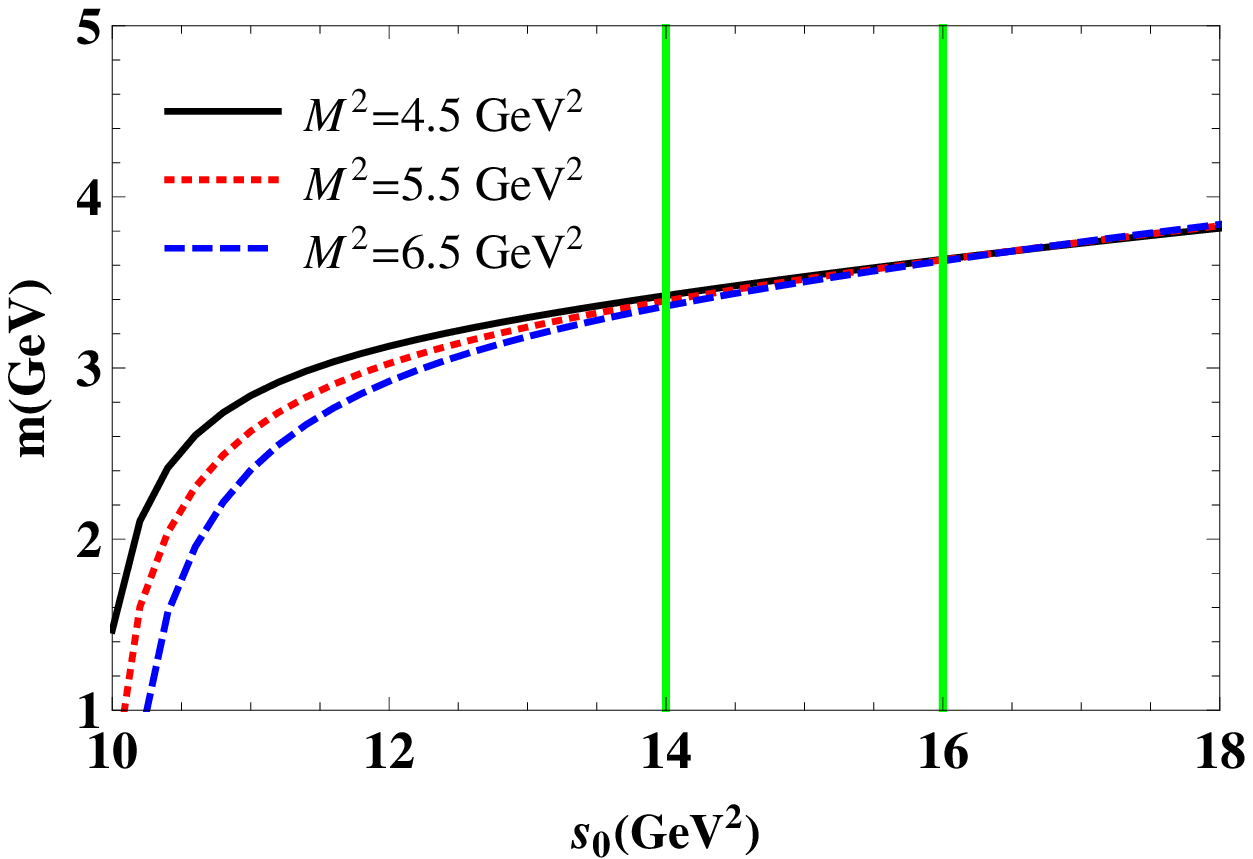}
\end{center}
\caption{Dependence of the $Z_{\mathrm{V}}$ tetraquark's mass $m$ on the Borel parameter $M^{2}$ (left panel), and on the continuum threshold parameter $s_0$ (right panel). Vertical lines fix boundaries of working regions for $M^{2}$ and $s_0$ used in numerical computations.}
\label{fig:Mass}
\end{figure}
\begin{figure}[h!]
\begin{center}
\includegraphics[totalheight=6cm,width=8cm]{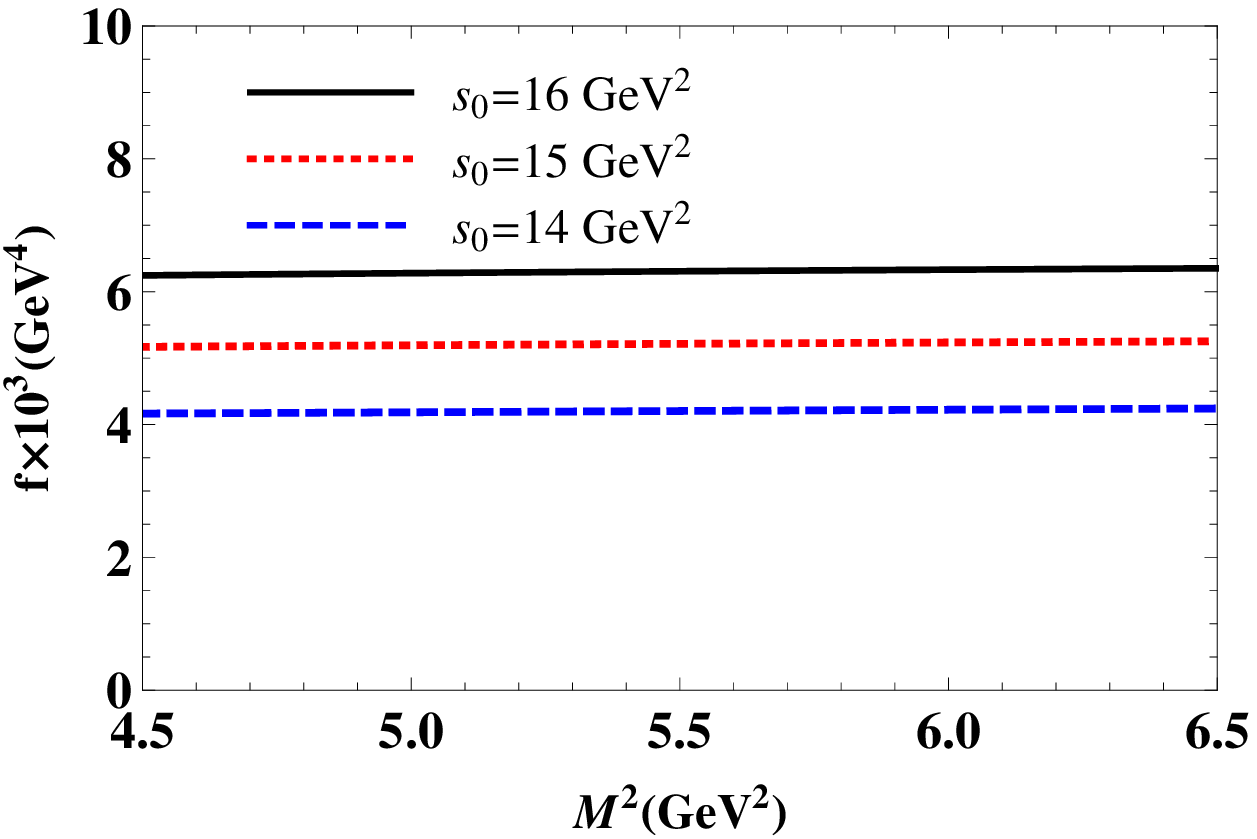}
\includegraphics[totalheight=6cm,width=8cm]{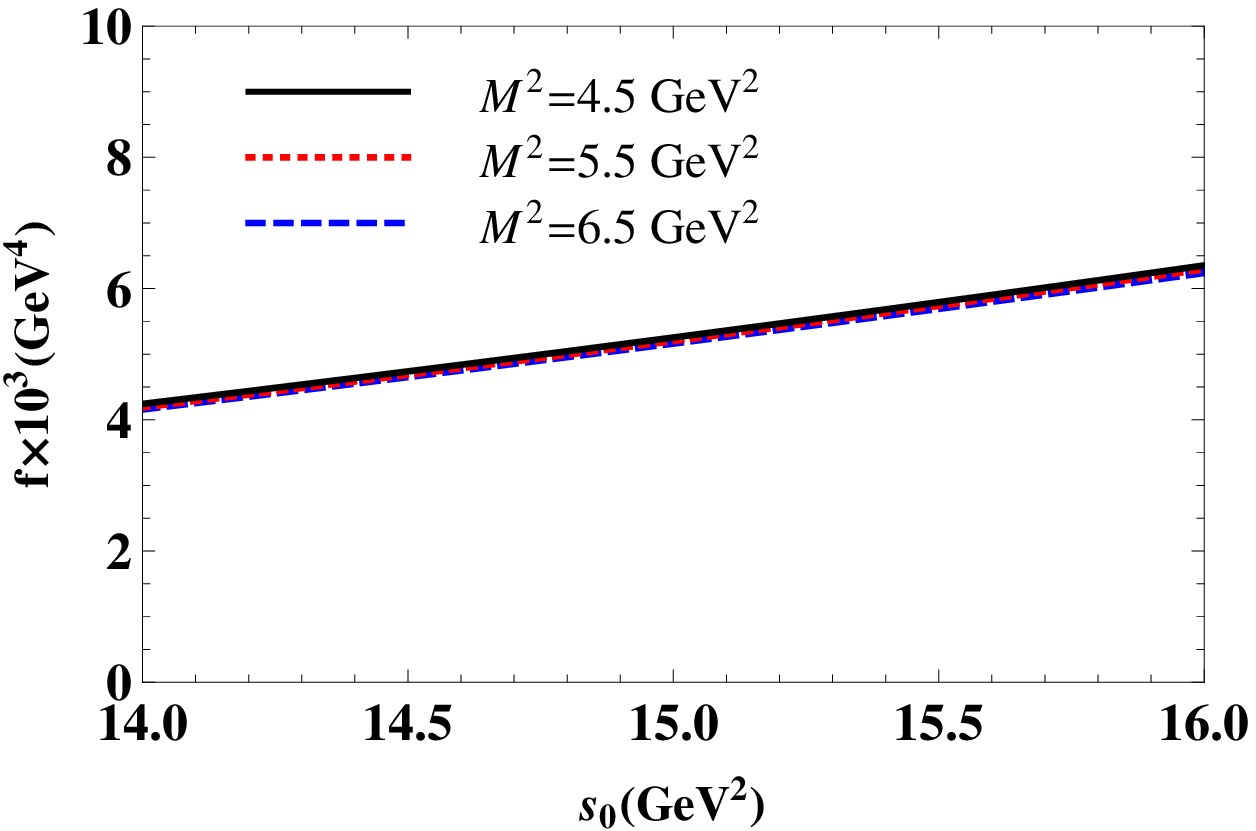}
\end{center}
\caption{The coupling $f$ of the tetraquark $Z_{\mathrm{V}}$ as a function of $M^{2}$ (left panel), and $s_0$ (right panel). Variations of $f$ are shown within working limits of the parameters $M^{2}$ and $s_0$.}
\label{fig:Coupling}
\end{figure}

\end{widetext}

The second source of theoretical errors is connected with a choice of the
continuum threshold parameter $s_{0}$. Its working window should also
satisfy limits stemming from dominance of $\mathrm{PC}$ and convergence of $%
\mathrm{OPE}$. Despite $M^{2}$, $s_{0}$ bears physical information about
first excitation of the tetraquark $Z_{\mathrm{V}}$. The self-consistent
analysis implies that $\sqrt{s_{0}}$ has to be smaller than mass of such
state. There are only a few samples when two observed resonances were
assumed to be ground and radially excited states of the same tetraquark. The
resonances $Z_{c}(3900)$ and $Z_{c}(4430)$ may form one of such pairs \cite%
{Agaev:2017tzv}. The difference between masses of $Z_{c}(3900)$ and $%
Z_{c}(4430)$ is equal to $\approx 530~\mathrm{MeV}$, therefore a mass gap $%
\sqrt{s_{0}}-m\approx (400-600)~\mathrm{MeV}$ can be considered as a
reasonable estimate for tetraquarks. Here, we get on average $\sqrt{s_{0}}%
-m\approx 400~\mathrm{MeV}$ which is in accord with this general analysis.

Effects connected with a choice of parameters $M^{2}$ and $s_{0}$ are two
main sources of theoretical uncertainties in sum rule computations. In the
case of the mass $m$ these ambiguities equal to $\pm 3.6\%$, whereas for the
coupling $f$ \ they are $\pm 21\%$ of the full result. Theoretical
uncertainties for $f$ are larger than for the mass, nevertheless, they do
not overshoot accepted limits.

It is interesting to analyze a gap between masses of axial-vector and vector
tetraquarks with the same content. Our present calculations demonstrate that
for tetraquarks $[cu][\overline{s}\overline{d}]$ the difference between
masses of the axial-vector and vector particles $Z_{\mathrm{AV}}$ and $Z_{%
\mathrm{V}}$ is $m-m_{\mathrm{Z}_{\mathrm{AV}}}\approx 690~\mathrm{MeV}$.
This result can be compared with similar predictions for other tetraquarks.
Thus, resonances $Y(4660)$ and $X(4140)$ with spin-parities $J^{\mathrm{PC}%
}=1^{--}$ and $1^{++}$, quark content $[cs][\overline{c}\overline{s}]$ and
color structure $[\overline{\mathbf{3}}_{c}]\otimes \lbrack \mathbf{3}_{c}]$
were investigated in Refs.\ \cite{Sundu:2018toi,Agaev:2017foq},
respectively. An estimate for $m_{Y}-m_{X_{1}}$ extracted from these studies
is approximately equal to $500~\mathrm{MeV}$. In other words, radially
excited axial-vector and ground-state vector tetraquarks, in this special
case, are close in mass. This fact maybe useful to explain numerous charged
and neutral resonances from $XYZ$ family in the mass range of $4-5~\mathrm{%
GeV}$.

Another issue to be addressed here, is a mass gap between vector tetraquarks
$Z_{\mathrm{V}}$ and $X_{1}(2900)$, which amounts to $\approx 600~\mathrm{MeV%
}$ and is quite large. It has been noted in section \ref{sec:Int}, that $%
X_{1}$ can be modeled as a vector tetraquark $[ud][\overline{c}\overline{s}]$%
, and hence both $Z_{\mathrm{V}}$ and $X_{1}$ are composed of same quarks.
But there are two reasons, which may lead to aforementioned mass effect.
First of them is internal organizations of these states: The tetraquark $Z_{%
\mathrm{V}}$ is composed of a heavy diquark $[cu]$ and relatively heavy
antidiquark $[\overline{s}\overline{d}]$, whereas $X_{1}$ is strongly
heavy-light compound. The latter is more tightly connected structure and
should be lighter than $Z_{\mathrm{V}}$. Besides, diquarks in $Z_{\mathrm{V}%
} $ have fractional positive electric charges which generate repulsive
forces between them. In the case of $X_{1}$, between $[ud]$ and $[\overline{c%
}\overline{s}]$ exists attractive electromagnetic interaction. Whether these
features of $Z_{\mathrm{V}}$ and $X_{1}$ are enough to explain a mass gap
between them or there are other sources of this effect, requires additional
studies.


\section{Strong decays of the tetraquark $Z_{\mathrm{V}}$}

\label{sec:Decays}

The sum of partial widths of $Z_{\mathrm{V}}$ tetraquark's different decay
channels constitutes its full width. The result for $m$ obtained in the
previous section is necessary to fix kinematically allowed strong decay
modes of $Z_{\mathrm{V}}$. Decays to final states $D_{s1}(2460)\pi $ and $%
D_{s0}^{\ast }(2317)\rho $ (below, simply $D_{s1}\pi $ and $D_{s0}^{\ast
}\rho $) are among allowed $S$-wave process for the tetraquark $Z_{\mathrm{V}%
}$. \ The $P$-wave processes, which will be explored, are decays $Z_{\mathrm{%
V}}\rightarrow DK$ and $Z_{\mathrm{V}}\rightarrow D_{s}\pi $.


\subsection{Processes $Z_{\mathrm{V}}\rightarrow D_{s1}\protect\pi $ and $Z_{%
\mathrm{V}}\rightarrow D_{s0}^{\ast }\protect\rho $}


Here, we study the decays $Z_{\mathrm{V}}\rightarrow D_{s1}\pi $ and $Z_{%
\mathrm{V}}\rightarrow D_{s0}^{\ast }\rho $, and compute their partial
widths. We provide details of calculations for the first process and write
down only essential formulas and final results for the second decay. It
should be noted that the mass $m=(3515~\pm 125)~\mathrm{MeV}$ makes possible
$S$-wave decays of $Z_{\mathrm{V}}$ to final states $DK_{1}(1270),$ $%
D_{s}b_{1}(1235)$ and $D_{s}a_{1}(1260)$ as well. But widths of these
processes, as it will be explained below, are suppressed relative to
aforementioned two decays due to kinematical factors. Therefore, we restrict
ourselves by investigation of two dominant $S$-wave decays.

The width of the process $Z_{\mathrm{V}}\rightarrow D_{s1}\pi $ can be found
using the coupling $g_{1}$ that describes strong interactions of particles $%
Z_{\mathrm{V}}$,$\ D_{s1}$, and $\pi $ at the vertex $Z_{\mathrm{V}%
}D_{s1}\pi $. In order to evaluate $g_{1}$, we use the QCD light-cone sum
rule method \cite{Balitsky:1989ry} and a soft-meson approximation \cite%
{Belyaev:1994zk,Ioffe:1983ju}.

Starting point in LCSR method is the correlation function
\begin{equation}
\Pi _{\mu \nu }(p,q)=i\int d^{4}xe^{ipx}\langle \pi (q)|\mathcal{T}\{J_{\mu
}^{D_{1}}(x)J_{\nu }^{\dag }(0)\}|0\rangle ,  \label{eq:CorrF3}
\end{equation}%
where by $D_{1}$ in the current $J_{\mu }^{D_{1}}$ we denote the meson $%
D_{s1}$. The current $J_{\nu }(0)$ in the correlation function $\Pi _{\mu
\nu }(p,q)$ is introduced in Eq.\ (\ref{eq:CR1}). As interpolating current $%
J_{\mu }^{D_{1}}(x)$ for the axial-vector meson $D_{s1}$, we use the
expression
\begin{equation}
J_{\mu }^{D_{1}}(x)=\overline{s}_{l}(x)i\gamma _{5}\gamma _{\mu }c_{l}(x),
\label{eq:Dcur}
\end{equation}%
with $l$ being the color index.

The function $\Pi _{\mu \nu }(p,q)$ has to be rewritten in terms of the
physical parameters of the initial and final particles involved into the
decay. By taking into account the ground states in the $D_{s1}$ and $Z_{%
\mathrm{V}}$ channels, we get
\begin{eqnarray}
&&\Pi _{\mu \nu }^{\mathrm{Phys}}(p,q)=\frac{\langle 0|J_{\mu
}^{D_{1}}|D_{s1}\left( p\right) \rangle }{p^{2}-m_{1}^{2}}\langle
D_{s1}\left( p\right) \pi (q)|Z_{\mathrm{V}}(p^{\prime })\rangle  \notag \\
&&\times \frac{\langle Z_{\mathrm{V}}(p^{\prime })|J_{\nu }^{\dagger
}|0\rangle }{p^{\prime 2}-m^{2}}+\cdots,  \label{eq:CorrF4}
\end{eqnarray}%
where $p$, $q$ and $p^{\prime }=p+q$ are the momenta of $D_{s1}$, $\pi $ ,
and $Z_{\mathrm{V}}$, respectively. In Eq.\ (\ref{eq:CorrF4}) $m_{1}$ is the
mass of the meson $D_{s1}$, and the ellipses stand for contributions of
higher resonances and continuum states in the $D_{s1}$ and $Z_{\mathrm{V}}$
channels.

To get more detailed expression for $\Pi _{\mu \nu }^{\mathrm{Phys}}(p,q)$,
we utilize the matrix elements
\begin{equation}
\langle 0|J_{\mu }^{D_{1}}|D_{s1}\rangle =f_{1}m_{1}\epsilon _{\mu },\
\langle Z_{\mathrm{V}}(p^{\prime })|J_{\nu }^{\dagger }|0\rangle =fm\epsilon
_{\nu }^{\prime \ast },  \label{eq:Mel}
\end{equation}%
and model the vertex $\langle D_{s1}\left( p\right) K(q)|Z_{\mathrm{V}%
}(p^{\prime })\rangle $ in the following form%
\begin{eqnarray}
&&\langle D_{s1}\left( p\right) \pi (q)|Z_{\mathrm{V}}(p^{\prime })\rangle
=g_{1}\left[ \left( p\cdot p^{\prime }\right) \left( \epsilon ^{\ast }\cdot
\epsilon ^{\prime }\right) \right.  \notag \\
&&\left. -\left( p\cdot \epsilon ^{\prime }\right) \left( p^{\prime }\cdot
\epsilon ^{\ast }\right) \right] .  \label{eq:Vert}
\end{eqnarray}%
In Eqs.\ (\ref{eq:Mel}) and (\ref{eq:Vert}) $f_{1}$ and $\epsilon _{\mu }$
are the decay constant and polarization vector of the meson $D_{s1}$,
respectively: Polarization vector of the tetraquark $Z_{\mathrm{V}}$ in this
section is denoted by $\epsilon _{\nu }^{\prime }$. Using these matrix
elements in Eq.\ (\ref{eq:CorrF4}), it is not difficult to find
\begin{eqnarray}
&&\Pi _{\mu \nu }^{\mathrm{Phys}}(p,q)=g_{1}\frac{m_{1}f_{1}mf}{\left(
p^{2}-m_{1}^{2}\right) \left( p^{\prime 2}-m^{2}\right) }  \notag \\
&&\times \left( \frac{m^{2}+m_{1}^{2}}{2}g_{\mu \nu }-p_{\mu }p_{\nu
}^{\prime }\right) .  \label{eq:CorrF5}
\end{eqnarray}%
The function $\Pi _{\mu \nu }^{\mathrm{Phys}}(p,q)$ contains two structures
proportional to $g_{\mu \nu }$ and $p_{\mu }p_{\nu }^{\prime }$,
respectively. These structures and relevant invariant amplitudes can be
employed to extract the sum rule for the strong coupling $g_{1}$. We choose
to work with the term $\sim g_{\mu \nu }$ and corresponding invariant
amplitude.

The QCD side of the sum rule can be obtained using explicit expressions of
the correlation function $\Pi _{\mu \nu }(p,q)$ and interpolating currents.
Having contracted quark and antiquark fields in the correlation function, we
get
\begin{eqnarray}
&&\Pi _{\mu \nu }^{\mathrm{OPE}}(p,q)=\int d^{4}xe^{ipx}\varepsilon
\widetilde{\varepsilon }\left[ \gamma _{5}\widetilde{S}_{c}^{lc}(x){}\gamma
_{\mu }\gamma _{5}\right.  \notag \\
&&\left. \times \widetilde{S}_{s}^{ml}(-x){}{}\gamma _{\nu }\gamma _{5}%
\right] _{\alpha \beta }\langle \pi (q)|\overline{u}_{\alpha
}^{b}(0)d_{\beta }^{n}(0)|0\rangle ,  \label{eq:CorrF6}
\end{eqnarray}%
where $\alpha $ and $\beta $ are the spinor indexes.

The function $\Pi ^{\mathrm{OPE}}(p,q)$ contains local matrix elements of
the quark operator $\overline{u}d$ sandwiched between the vacuum and pion.
Simple transformations allow one to express $\langle \pi (q)|\overline{u}%
_{\alpha }^{b}(0)d_{\beta }^{n}(0)|0\rangle $ in terms of the pion's known
local matrix elements. To this end, we expand $\overline{u}(0)d(0)$ by
employing full set of Dirac matrices $\Gamma ^{J}=\mathbf{1},\ \gamma _{5},\
\gamma _{\mu },\ i\gamma _{5}\gamma _{\mu },\ \sigma _{\mu \nu }/\sqrt{2}$,
and project obtained operators onto the color-singlet states. These
manipulations lead to replacement
\begin{equation}
\overline{u}_{\alpha }^{b}(0)d_{\beta }^{n}(0)\rightarrow \frac{1}{12}\delta
^{bn}\Gamma _{\beta \alpha }^{J}\left[ \overline{u}(0)\Gamma ^{J}d(0)\right]
,  \label{eq:MatEx}
\end{equation}%
which should be fulfilled in $\Pi _{\mu \nu }^{\mathrm{OPE}}(p,q)$. The
operators $\overline{u}(0)\Gamma ^{J}d(0)$ placed between the vacuum and
pion are matrix elements of the $\pi $ meson.

The QCD expression of the correlation function (\ref{eq:CorrF6}) contains
hard-scattering and soft parts. The hard-scattering part of $\Pi _{\mu \nu
}^{\mathrm{OPE}}(p,q)$ is expressed in terms of quark propagators, whereas
matrix elements of the pion form its soft component. For vertices built of
conventional mesons correlation functions depend on non-local matrix
elements of a final meson which are expressible in terms of its distribution
amplitudes (DAs). In the case under analysis, due to four-quark nature of $%
Z_{\mathrm{V}}$, $\Pi _{\mu \nu }^{\mathrm{OPE}}(p,q)$ contains only local
matrix elements of the pion. As a result, integrals over DAs which are
typical for LCSR method, reduce to overall normalization factors. In this
case, the correlation function has to be found by means of the soft-meson
approximation which implies computation of the hard-scattering part of $\Pi
_{\mu \nu }^{\mathrm{OPE}}(p,q)$ in the limit $q\rightarrow 0$ \cite%
{Agaev:2016dev}. It is worth to emphasize that soft-meson approximation is
necessary to analyze tetraquark-meson-meson vertices: Strong couplings at
vertices composed of two tetraquarks and a meson can be calculated by
employing full version of the LCSR method \cite{Agaev:2016srl}.

It is evident, that soft approximation should by applied also to the
phenomenological side of the sum rule. In the limit $q\rightarrow 0$ for the
amplitude $\Pi ^{\mathrm{Phys}}(p^{2})$, we get
\begin{equation}
\Pi ^{\mathrm{Phys}}(p^{2})=g_{1}\frac{f_{1}m_{1}fm}{(p^{2}-\widetilde{m}%
^{2})^{2}}\widetilde{m}^{2}+\cdots ,  \label{eq:CF2a}
\end{equation}%
where $\widetilde{m}^{2}=(m^{2}+m_{1}^{2})/2$. The function $\Pi ^{\mathrm{%
Phys}}(p^{2})$ depends on one variable $p^{2}=p^{\prime 2}$ and has the
double pole at $p^{2}=\widetilde{m}^{2}$. The Borel transformation of $\Pi ^{%
\mathrm{Phys}}(p^{2})$ is given by the formula
\begin{equation}
\Pi ^{\mathrm{Phys}}(M^{2})=g_{1}f_{1}m_{1}fm\widetilde{m}^{2}\frac{e^{-%
\widetilde{m}^{2}/M^{2}}}{M^{2}}+\cdots .  \label{eq:CF3a}
\end{equation}

Besides ground-state contribution, the amplitude $\Pi ^{\mathrm{Phys}%
}(M^{2}) $ in the soft limit contains unsuppressed terms which survive even
after Borel transformation. These contaminations should be removed from $\Pi
^{\mathrm{Phys}}(M^{2})$ by applying the operator \cite%
{Belyaev:1994zk,Ioffe:1983ju}
\begin{equation}
\mathcal{P}(M^{2},\widetilde{m}^{2})=\left( 1-M^{2}\frac{d}{dM^{2}}\right)
M^{2}e^{\widetilde{m}^{2}/M^{2}}.  \label{eq:Oper}
\end{equation}%
After this operation, remaining undesired terms in $\Pi ^{\mathrm{Phys}%
}(M^{2})$ can be subtracted by usual manner in the context of quark-hadron
duality assumption. It is clear, that we have to apply the operator $%
\mathcal{P}(M^{2},\widetilde{m}^{2})$ to QCD side of the sum rule as well.
Then, the sum rule for the strong coupling $g_{1}$ reads
\begin{equation}
g_{1}=\frac{1}{f_{1}m_{1}fm\widetilde{m}^{2}}\mathcal{P}(M^{2},\widetilde{m}%
^{2})\Pi ^{\mathrm{OPE}}(M^{2},s_{0}),  \label{eq:SRcoupl}
\end{equation}%
where $\Pi ^{\mathrm{OPE}}(M^{2},s_{0})$ is Borel transformed and subtracted
invariant amplitude $\Pi ^{\mathrm{OPE}}(p^{2})$ corresponding to the
structure $g_{\mu \nu }$ in $\Pi _{\mu \nu }^{\mathrm{OPE}}(p,q)$.

Prescriptions to calculate the correlation function $\Pi _{\mu \nu }^{%
\mathrm{OPE}}(p,q)$ in the soft approximation were presented in Ref.\ \cite%
{Agaev:2016dev}, and in many other publications, for this reason we do not
concentrate here on details. Computations of $\Pi ^{\mathrm{OPE}%
}(M^{2},s_{0})$ performed in accordance with this scheme lead to the
expression
\begin{eqnarray}
&&\Pi ^{\mathrm{OPE}}(M^{2},s_{0})=\frac{f_{\pi }\mu _{\pi }}{48\pi ^{2}}%
\int_{\mathcal{M}^{2}}^{s_{0}}\frac{ds(m_{c}^{2}-s)}{s^{2}}  \notag \\
&&\times \left( m_{c}^{4}+m_{c}^{2}s+6m_{c}m_{s}s-2s^{2}\right) e^{-s/M^{2}}
\notag \\
&&+\Pi _{\mathrm{NP}}(M^{2}),  \label{eq:DecayCF}
\end{eqnarray}%
where the first term is the perturbative contribution to $\Pi ^{\mathrm{OPE}%
}(M^{2},s_{0})$. The nonperturbative component $\Pi _{\mathrm{NP}}(M^{2})$
of the correlation function is calculated with dimension-$9$ accuracy, and
has the following form%
\begin{eqnarray}
&&\Pi _{\mathrm{NP}}(M^{2})=\frac{\langle \overline{s}s\rangle f_{\pi }\mu
_{\pi }m_{c}\left( 2M^{2}+m_{c}m_{s}\right) }{12M^{2}}e^{-m_{c}^{2}/M^{2}}
\notag \\
&&+\langle \frac{\alpha _{s}G^{2}}{\pi }\rangle \frac{f_{\pi }\mu _{\pi
}m_{c}}{144M^{4}}\int_{0}^{1}\frac{dx}{X^{3}}\left[
m_{c}^{2}m_{s}X-2m_{s}XM^{2}\right.  \notag \\
&&\left. +m_{c}^{3}x+m_{c}xXM^{2}\right] e^{m_{c}^{2}/[M^{2}X]}+\frac{%
\langle \overline{s}g\sigma Gs\rangle f_{\pi }\mu _{\pi }}{72M^{6}}  \notag
\\
&&\times \left(
m_{s}M^{4}+m_{s}m_{c}^{2}M^{2}-3m_{c}^{3}M^{2}-m_{s}m_{c}^{4}\right)
e^{-m_{c}^{2}/M^{2}}  \notag \\
&&+\langle \frac{\alpha _{s}G^{2}}{\pi }\rangle \langle \overline{s}s\rangle
\frac{f_{\pi }\mu _{\pi }\pi ^{2}m_{c}(3M^{2}-m_{c}^{2})}{216M^{8}}  \notag
\\
&&\times \left( 2M^{2}+m_{s}m_{c}\right) e^{-m_{c}^{2}/M^{2}}+\langle \frac{%
\alpha _{s}G^{2}}{\pi }\rangle \langle \overline{s}g\sigma Gs\rangle  \notag
\\
&&\times \frac{f_{\pi }\mu _{\pi }\pi ^{2}m_{c}\left(
m_{c}^{5}m_{s}+3M^{2}m_{c}^{4}-7m_{c}^{3}m_{s}M^{2}\right. }{1296M^{12}}
\notag \\
&&\left. -18m_{c}^{2}M^{4}+8m_{c}m_{s}M^{4}+18M^{6}\right)
e^{-m_{c}^{2}/M^{2}},  \label{eq:DecayNPCF}
\end{eqnarray}%
where $X=x-1$.

Contributions to $\Pi _{\mathrm{NP}}(M^{2})$ arise from the matrix element
\begin{equation}
\langle 0|\overline{d}i\gamma _{5}u|\pi \rangle =f_{\pi }\mu _{\pi },
\end{equation}%
where
\begin{equation}
\mu _{\pi }=\frac{m_{\pi }^{2}}{m_{u}+m_{d}}=-\frac{2\langle \overline{q}%
q\rangle }{f_{\pi }^{2}}.  \label{eq:MEpion}
\end{equation}%
The last equality in Eq.\ (\ref{eq:MEpion}) is the relation between the mass
$m_{\pi }$ and decay constant $f_{\pi }$ of the pion, masses of quarks, and
quark condensate $\langle \overline{q}q\rangle $, which stems from the
partial conservation of the axial-vector current.

\begin{table}[tbp]
\begin{tabular}{|c|c|}
\hline\hline
Parameters & Values (in $\mathrm{MeV}$ units) \\ \hline\hline
$m_1=m[D_{s1}(2460)]$ & $2459.5\pm 0.6$ \\
$f_1=f[D_{s1}(2460)]$ & $225 \pm 25$ \\
$m_2=m[D_{s0}^{*}(2317)]$ & $2317.8\pm 0.5$ \\
$f_2=f[D_{s0}^{*}(2317)]$ & $202 \pm 15$ \\
$m_{D}$ & $1869.65\pm 0.05$ \\
$f_{D}$ & $212.6 \pm 0.7$ \\
$m_{D_s}$ & $1968.34\pm 0.07$ \\
$f_{D_s}$ & $249.9 \pm 0.5$ \\
$m_{K}$ & $493.677\pm 0.0016$ \\
$f_{K}$ & $155.7 \pm 0.3$ \\
$m_{\pi}$ & $139.57039 \pm 0.00018$ \\
$f_{\pi}$ & $130.2 \pm 1.2 $ \\
$m_{\rho}$ & $775.26\pm 0.25$ \\
$f_{\rho}$ & $216 \pm 3$ \\ \hline\hline
\end{tabular}%
\caption{Masses and decay constants of mesons used in numerical analyses.}
\label{tab:Param}
\end{table}

In numerical computations of $\Pi ^{\mathrm{OPE}}(M^{2},s_{0})$, we choose $%
M^{2}$ and $s_{0}$ within limits given by Eq.\ (\ref{eq:Wind1}). Besides $%
M^{2}$ and $s_{0}$, Eq.\ (\ref{eq:SRcoupl}) contains various vacuum
condensates and spectroscopic parameters of the final-state mesons $D_{s1}$
and $\pi $. The masses and decay constants of the mesons $D_{s1}$ and $\pi $%
, as well as other parameters used in numerical analyses are borrowed from
Ref.\ \cite{PDG:2020} and presented in Table\ \ref{tab:Param}: For decay
constants of the mesons $D_{s0}^{\ast }$ and $D_{s1}$, we utilize the sum
rule predictions from Refs.\ \cite{Narison:2015nxh} and \cite%
{Colangelo:2005hv}, respectively.

For the coupling $g_{1}$, we get
\begin{equation}
g_{1}=0.36_{-0.06}^{+0.09}~\mathrm{GeV}^{-1}  \label{eq:Coupl1}
\end{equation}%
The partial width of the decay $Z_{\mathrm{V}}\rightarrow D_{s1}\pi $ can be
found by means of the formula
\begin{equation}
\Gamma _{1}\left[ Z_{\mathrm{V}}\rightarrow D_{s1}\pi \right] =\frac{%
g_{1}^{2}m_{1}^{2}\lambda }{24\pi }\left( 3+\frac{2\lambda ^{2}}{m_{1}^{2}}%
\right) ,  \label{eq:DW}
\end{equation}%
where $\lambda \equiv \lambda \left( m,m_{1},m_{\pi }\right) $ and
\begin{eqnarray}
\lambda \left( a,b,c\right) &=&\frac{1}{2a}\left[ a^{4}+b^{4}+c^{4}\right.
\notag \\
&&\left. -2\left( a^{2}b^{2}+a^{2}c^{2}+b^{2}c^{2}\right) \right] ^{1/2}.
\end{eqnarray}%
By employing this expression, it is not difficult to obtain
\begin{equation}
\Gamma _{1}\left[ Z_{\mathrm{V}}\rightarrow D_{s1}\pi \right]
=29.8_{-9.4}^{+17.6}~\mathrm{MeV}.  \label{eq:DW1Numeric}
\end{equation}

The strong coupling and partial width of the second process $Z_{\mathrm{V}%
}\rightarrow D_{s0}^{\ast }\rho $ can be found by the same manner. Here, one
starts from the correlation function%
\begin{equation}
\widetilde{\Pi }_{\nu }(p,q)=i\int d^{4}xe^{ipx}\langle \rho (q)|\mathcal{T}%
\{J^{D_{0}}(x)J_{\nu }^{\dag }(0)\}|0\rangle ,  \label{eq:CorrF3a}
\end{equation}%
where the interpolating current of the scalar meson $D_{s0}^{\ast }$ is
denoted by $J^{D_{0}}(x)$ and determined by the expression%
\begin{equation}
J^{D_{0}}(x)=\overline{s}_{l}(x)c_{l}(x).
\end{equation}%
In our studies, we use the matrix element $\langle 0|J^{D_{0}}|D_{s0}^{\ast
}\rangle =f_{2}m_{2}$, with $m_{2}$ and $f_{2}$ being the mass and decay
constant of the $D_{s0}^{\ast }$. The vertex $Z_{\mathrm{V}}D_{s0}^{\ast
}\rho $ is modeled in the form
\begin{eqnarray}
&&\langle D_{s0}^{\ast }\left( p\right) \rho (q)|Z_{\mathrm{V}}(p^{\prime
})\rangle =g_{2}\left[ \left( q\cdot p^{\prime }\right) \left( \epsilon
^{\ast }\cdot \epsilon ^{\prime }\right) \right.  \notag \\
&&\left. -\left( q\cdot \epsilon ^{\prime }\right) \left( p^{\prime }\cdot
\epsilon ^{\ast }\right) \right] ,
\end{eqnarray}%
where $\epsilon _{\mu }$ is the polarization vector of the $\rho $ meson.
Then, the phenomenological and QCD sides of the sum rule are given by
expressions%
\begin{eqnarray}
&&\widetilde{\Pi }_{\nu }^{\mathrm{Phys}}(p,q)=g_{2}\frac{m_{2}f_{2}mf}{%
\left( p^{2}-m_{2}^{2}\right) \left( p^{\prime 2}-m^{2}\right) }  \notag \\
&&\times \left( \frac{m_{2}^{2}-m^{2}}{2}\epsilon _{\nu }^{\ast }+p^{\prime
}\cdot \epsilon ^{\ast }q_{\nu }\right) ,
\end{eqnarray}%
and%
\begin{eqnarray}
&&\widetilde{\Pi }_{\nu }^{\mathrm{OPE}}(p,q)=-i\int
d^{4}xe^{ipx}\varepsilon \widetilde{\varepsilon }\left[ \gamma _{5}%
\widetilde{S}_{c}^{lc}(x){}\widetilde{S}_{s}^{ml}(-x)\right.  \notag \\
&&\left. \times {}{}\gamma _{\nu }\gamma _{5}\right] _{\alpha \beta }\langle
\rho (q)|\overline{u}_{\alpha }^{b}(0)d_{\beta }^{n}(0)|0\rangle ,
\end{eqnarray}%
respectively. We perform calculations using the structures $\sim \epsilon
_{\nu }^{\ast }$ in $\widetilde{\Pi }_{\nu }^{\mathrm{Phys}}$ and $%
\widetilde{\Pi }_{\nu }^{\mathrm{OPE}}$. The relevant amplitude $\widetilde{%
\Pi }^{\mathrm{OPE}}(M^{2},s_{0})$ receives contributions from the matrix
elements of the $\rho $ meson \
\begin{eqnarray}
\langle 0|\overline{d}\gamma _{\mu }u|\rho \rangle &=&f_{\rho }m_{\rho
}\epsilon _{\mu },  \notag \\
\langle 0|\overline{d}g\widetilde{G}_{\mu \nu }\gamma _{\nu }\gamma
_{5}u|\rho \rangle &=&f_{\rho }m_{\rho }^{3}\zeta _{4}\epsilon _{\mu },
\label{eq:MERho}
\end{eqnarray}%
where $m_{\rho }$ and $f_{\rho }$ are the mass and decay constant of the $%
\rho $ meson, and $\widetilde{G}_{\mu \nu }$ is the dual gluon field tensor $%
\widetilde{G}_{\mu \nu }=\varepsilon _{\mu \nu \alpha \beta }G^{\alpha \beta
}/2$. The second equality in Eq.\ (\ref{eq:MERho}) is the matrix element of
the twist-4 operator \cite{Ball:1998ff}. The parameter $\zeta _{4\rho }$ was
evaluated in the context of QCD sum rule approach at the renormalization
scale $\mu =1\,\,{\mathrm{GeV}}$ in Ref.\ \cite{Ball:2007zt} and is equal to
$\zeta _{4\rho }=0.07\pm 0.03$.

The correlation function $\widetilde{\Pi }^{\mathrm{OPE}}(M^{2},s_{0})$ has
the form
\begin{eqnarray}
&&\widetilde{\Pi }^{\mathrm{OPE}}(M^{2},s_{0})=\frac{f_{\rho }m_{\rho }}{%
16\pi ^{2}}\int_{\mathcal{M}^{2}}^{s_{0}}\frac{ds(s-m_{c}^{2})}{s}  \notag \\
&&\times \left( m_{c}^{2}+2m_{c}m_{s}-s\right) e^{-s/M^{2}}+\frac{f_{\rho
}m_{\rho }^{3}\zeta _{4\rho }}{32\pi ^{2}}  \notag \\
&&\times \int_{\mathcal{M}^{2}}^{s_{0}}\frac{ds(m_{c}^{4}-s^{2})}{s^{2}}%
e^{-s/M^{2}}+\widetilde{\Pi }_{\mathrm{NP}}(M^{2}),  \label{eq:DecayCF1}
\end{eqnarray}%
where $\widetilde{\Pi }_{\mathrm{NP}}(M^{2})$ is nonperturbative component
of $\widetilde{\Pi }^{\mathrm{OPE}}$: We refrain from providing its explicit
expression here.

Omitting further details, we write down results for the strong coupling $%
g_{2}$ and partial width of the process $Z_{\mathrm{V}}\rightarrow
D_{s0}^{\ast }\rho $
\begin{eqnarray}
&&g_{2}=0.61_{-0.12}^{+0.18}~\mathrm{GeV}^{-1},  \notag \\
&&\Gamma _{2}\left[ Z_{\mathrm{V}}\rightarrow D_{s0}^{\ast }\rho \right]
=9.3_{-3.1}^{+6.5}~\mathrm{MeV}.  \label{eq:DWA}
\end{eqnarray}%
Returning to other $S$-wave decays listed above, we assume that couplings
corresponding to vertices $Z_{\mathrm{V}}DK_{1}(1270)$ etc., are the same
order of $g_{1}$ and $g_{2}$. Then widths of these decays are suppressed,
because the factor $m_{\ast }^{2}$$\lambda \left( 3+2\lambda ^{2}/m_{\ast
}^{2}\right) $ ($m_{\ast }$ is a mass of a final meson) is smaller for two
final-state mesons of approximately equal mass than in the case of light and
heavy mesons. We also take into account that the full width $\Gamma _{%
\mathrm{full}}$ of the tetraquark $Z_{\mathrm{V}}$ is formed mainly due to $%
P $-wave processes $Z_{\mathrm{V}}\rightarrow DK$ and $Z_{\mathrm{V}%
}\rightarrow D_{s}\pi $, and for this reason consider only two dominant $S$%
-wave decays.


\subsection{Decays $Z_{\mathrm{V}}\rightarrow DK$ and $Z_{\mathrm{V}%
}\rightarrow D_{s}\protect\pi $}


In this subsection, we consider the $P$-wave decays $Z_{\mathrm{V}%
}\rightarrow DK$ and $Z_{\mathrm{V}}\rightarrow D_{s}\pi $ of the tetraquark
$Z_{\mathrm{V}}$. Treatments of these processes do not differ from analysis
carried out above, differences being mainly in meson-tetraquark vertices and
matrix elements of final-state mesons employed in calculations.

Let us concentrate on the decay $Z_{\mathrm{V}}\rightarrow DK$. The
correlation function to find a sum rule for the strong coupling $G_{1}$ of
vertex $Z_{\mathrm{V}}DK$ is given by the formula
\begin{equation}
\Pi _{\nu }(p,q)=i\int d^{4}xe^{ipx}\langle K(q)|\mathcal{T}\{J^{D}(x)J_{\nu
}^{\dag }(0)\}|0\rangle ,  \label{eq:CorrF3b}
\end{equation}%
where $J^{D}(x)$ is the interpolating current

\begin{equation}
J^{D}(x)=\overline{d}_{l}(x)i\gamma _{5}c_{l}(x),  \label{eq:Curr}
\end{equation}%
for the pseudoscalar meson $D$.

Then, the physical side of the sum rule has the form%
\begin{eqnarray}
&&\Pi _{\nu }^{\mathrm{Phys}}(p,q)=\frac{\langle 0|J^{D}|D\left( p\right)
\rangle }{p^{2}-m_{D}^{2}}\langle D\left( p\right) K(q)|Z_{\mathrm{V}%
}(p^{\prime })\rangle  \notag \\
&&\times \frac{\langle Z_{\mathrm{V}}(p^{\prime })|J_{\nu }^{\dagger
}|0\rangle }{p^{\prime 2}-m^{2}}+\cdots.  \label{eq:CorrF4a}
\end{eqnarray}%
Using the matrix elements
\begin{equation}
\langle 0|J^{D}|D\rangle =\frac{f_{D}m_{D}^{2}}{m_{c}},\ \langle D\left(
p\right) K(q)|Z_{\mathrm{V}}(p^{\prime })\rangle =G_{1}p\cdot \epsilon
^{\prime },  \label{eq:MElD}
\end{equation}%
for $\Pi _{\nu }^{\mathrm{Phys}}$, we find
\begin{eqnarray}
&&\Pi _{\nu }^{\mathrm{Phys}}(p,q)=G_{1}\frac{f_{D}m_{D}^{2}fm}{%
2m_{c}(p^{2}-m_{D}^{2})(p^{\prime 2}-m^{2})}  \notag \\
&&\times \left[ \left( -1+\frac{m_{D}^{2}-m_{K}^{2}}{m^{2}}\right) p_{\nu
}+\left( 1+\frac{m_{D}^{2}-m_{K}^{2}}{m^{2}}\right) q_{\nu }\right]  \notag
\\
&&+\cdots.  \label{eq:CorrF4b}
\end{eqnarray}%
In expressions above, $m_{D}$ and $f_{D}$ are the mass and decay constant of
the $D$ meson, respectively.

In terms of quark propagators the same correlation function $\Pi _{\upsilon
}(p,q)$ is determined by the expression%
\begin{eqnarray}
&&\Pi _{\nu }^{\mathrm{OPE}}(p,q)=\int d^{4}xe^{ipx}\varepsilon \widetilde{%
\varepsilon }\left[ \gamma _{5}\widetilde{S}_{c}^{lc}(x){}\gamma _{5}%
\widetilde{S}_{d}^{nl}(-x)\right.  \notag \\
&&\left. \times \gamma _{5}{}{}\gamma _{\nu }\right] _{\alpha \beta }\langle
K(q)|\overline{u}_{\alpha }^{b}(0)s_{\beta }^{m}(0)|0\rangle .
\label{eq:CorrF6a}
\end{eqnarray}%
Operations necessary to derive the sum rule for the coupling $G_{1}$ have
just been explained above, that is why we do not consider these questions.
Let us note that the sum rule for $G_{1}$ has been obtained using the
structures proportional $p_{\nu }$. The local matrix element of $K$ meson
which contributes to $\Pi _{\nu }^{\mathrm{OPE}}(M^{2},s_{0})$ is
\begin{equation}
\langle 0|\overline{d}i\gamma _{5}u|\pi \rangle =\frac{f_{K}m_{K}^{2}}{m_{s}}%
,
\end{equation}%
where $m_{K}$ and $f_{K}$ are the mass and decay constant of the $K$ meson.

The width of the process $Z_{\mathrm{V}}\rightarrow DK$ can be found by
means of the formula
\begin{equation}
\Gamma _{3}\left[ Z_{\mathrm{V}}\rightarrow DK\right] =\frac{%
G_{1}^{2}\lambda ^{3}(m,m_{D,}m_{K})}{24\pi m^{2}}.
\end{equation}%
In sum rule computations of the coupling $G_{1}$ the Borel and continuum
threshold parameters are chosen as in Eq.\ (\ref{eq:Wind1}). The
spectroscopic parameters of the mesons $D$ and $K$ are collected in Table\ %
\ref{tab:Param}. Our predictions read%
\begin{eqnarray}
&&G_{1}=4.3_{-0.7}^{+1.2},  \notag \\
&&\Gamma _{3}\left[ Z_{\mathrm{V}}\rightarrow DK\right]
=34.6_{-10.9}^{+20.6}~\mathrm{MeV}.  \label{eq:DWB}
\end{eqnarray}%
For the second $P$-wave decay $Z_{\mathrm{V}}\rightarrow D_{s}\pi $, we find
\begin{eqnarray}
&&G_{2}=6.6_{-1.1}^{+1.8},  \notag \\
&&\Gamma _{4}\left[ Z_{\mathrm{V}}\rightarrow D_{s}\pi \right]
=81.8_{-25.8}^{+48.9}~\mathrm{MeV}.
\end{eqnarray}

Decay channels of the tetraquark $Z_{\mathrm{V}}$ considered in this section
allow us to evaluate its full width

\begin{equation}
\Gamma _{\mathrm{full}}=156_{-30}^{+56}~\mathrm{MeV}.
\end{equation}%
It is clear that $Z_{\mathrm{V}}$ can be classified as an exotic meson of
wide width. Its main decay modes are processes $Z_{\mathrm{V}}\rightarrow
D_{s}\pi $ and $Z_{\mathrm{V}}\rightarrow DK$ with branching ratios $%
\mathcal{BR}(Z_{\mathrm{V}}\rightarrow D_{s}\pi )\approx 0.52$ and $\mathcal{%
BR}(Z_{\mathrm{V}}\rightarrow DK)\approx 0.22$, respectively. Contribution
of the decay $Z_{\mathrm{V}}\rightarrow D_{s1}\pi $ is also considerable
with estimate $\mathcal{BR}(Z_{\mathrm{V}}\rightarrow D_{s}\pi )\approx 0.19$%
.


\section{Conclusions and final notes}

\label{sec:Disc}

In this article, we have studied the doubly charged vector tetraquark $Z_{%
\mathrm{V}}^{++}=[cu][\overline{s}\overline{d}]$ and calculated its mass $m$
and width $\Gamma _{\mathrm{full}}$. The parameters of $Z_{\mathrm{V}}^{++}$
have been evaluated using the QCD two-point and light-cone sum rules. The
doubly charged tetraquarks $[sd][\overline{u}\overline{c}]$ with
spin-parities $J^{\mathrm{P}}=0^{+}$, $0^{-}$ and $1^{+}$ were investigated
in our paper \cite{Agaev:2017oay}. The exotic mesons $Z^{++}$ are built of
four quarks of different flavors, and moreover bear two units of electric
charge. These particles were not found till now, but due to unique features
are interesting objects for both theoretical and experimental studies.

The LHCb collaboration recently observed structures $X_{0(1)}(2900)$ which
may be interpreted as exotic mesons containing four different quarks \cite%
{Aaij:2020hon,Aaij:2020ypa}. The structures $X_{0(1)}$ were seen as
resonance-like peaks in the mass distribution of $D^{-}K^{+}$ mesons. The
master process to discover $X_{0(1)}$ was the weak decay of the $B$ meson $%
B^{+}\rightarrow D^{+}D^{-}K^{+}$. It is remarkable that the process $%
B^{+}\rightarrow D^{+}D^{-}K^{+}$ and data collected during its exploration
can be employed to observe another tetraquarks, namely states with quark
content $cu\overline{s}\overline{d}$. In fact, analysis of the invariant
mass distribution of mesons $D^{+}K^{+}$ may lead to information about the
scalar and vector tetraquarks $Z_{\mathrm{S}}^{++}$ and $Z_{\mathrm{V}}^{++}$%
. Decays to final mesons $D^{+}K^{+}$ are among favored channels of these
particles. In fact, relevant branching ratios are equal to $\mathcal{BR}(Z_{%
\mathrm{S}}\rightarrow DK)\approx 0.86$ and $\mathcal{BR}(Z_{\mathrm{V}%
}\rightarrow DK)\approx 0.22$, respectively.

The decay of the $B$ meson $B^{+}\rightarrow D^{-}D_{s}^{+}\pi ^{+}$ is also
useful to see states $Z_{\mathrm{S}}^{++}$ and $Z_{\mathrm{V}}^{++}$,
because processes $Z_{\mathrm{S}}\rightarrow D_{s}\pi $ and $Z_{\mathrm{V}%
}\rightarrow D_{s}\pi $ have considerable branching ratios $\mathcal{BR}(Z_{%
\mathrm{S}}\rightarrow D_{s}\pi )\approx 0.12$ and $\mathcal{BR}(Z_{\mathrm{V%
}}\rightarrow D_{s}\pi )\approx 0.52$. The decay $B^{+}\rightarrow
D^{-}D_{s}^{+}\pi ^{+}$ can be used to discover also tetraquarks $[ud][%
\overline{c}\overline{d}]$: These neutral states may be fixed in the
invariant mass distribution of $D^{-}\pi ^{+}$ mesons.

We have modeled $Z_{\mathrm{V}}^{++}$ as a four-quark exotic meson with
diquark-antidiquark structure. Our analyses have proved that the
interpolating current (\ref{eq:CR1}) used in the framework of the QCD sum
rule method correctly describes the particle $Z_{\mathrm{V}}^{++}$ and leads
to reliable results for its parameters. In fact, existence of the working
windows for parameters $M^{2}$ and $s_{0}$ that satisfy standard
requirements of the sum rules, and self-consistency of used $\sqrt{s_{0}}$
and extracted $m$ argue in favor of this conclusion. It is known that a
diquark-antidiquarks are tightly bound states, and may be favorite forms for
doubly charged tetraquarks. But such four-quark systems may exist also as
hadronic molecules. Thus, doubly charged molecular compounds with the quark
content $\overline{Q}\overline{Q}qq$, where $Q$ is $c$ or $b$-quark were
considered in Ref. \cite{Ohkoda:2012hv}. In this article the authors used
the heavy quark effective theory to derive interactions between heavy
mesons, and coupled channel Schrodinger equations to find the bound and/or
resonant states with various quantum numbers. It was demonstrated that, for
example, $D$ and $D^{\ast }$ mesons can form doubly charged resonant state
with $I(J^{\mathrm{P}})=1(0^{-})$. Similar analysis of a molecule
counterpart of $Z_{\mathrm{V}}^{++}$ in the context of the QCD sum rule
method implies usage of molecular-type interpolating current. Whether such
current would lead to strong sum rule predictions or not requires detailed
analysis, which however is beyond the scope of the present paper.

It is seen, that three-meson weak decays of $B$ meson are sources of
valuable information on four-quark exotic states. Data collected by various
collaborations in running experiments can be utilized for these purposes.
New decays of $B$ meson can open wide prospects to study numerous four-quark
states. Indeed, final-state mesons in such processes can be combined to form
different pairs and their invariant mass distributions can be explored to
detect resonance-type enhancements. In any case, additional experimental and
theoretical studies are necessary to take advantage of emerging
opportunities.

\end{document}